\documentclass[runningheads]{llncs}
\usepackage[T1]{fontenc}
\usepackage{tabularx}
\usepackage{graphicx}
\usepackage{todonotes}

\usepackage{appendix}

\begin{document}
\title{Experiversum: an Ecosystem for Curating and Enhancing Data-Driven Experimental Science}%
%
\titlerunning{Curating and Enhancing Data-Driven Experimental Science}
%
\author{Genoveva Vargas-Solar\inst{1}
\and Umberto Costa\inst{2}
\and Jérôme Darmont\inst{3}
\and Javier A. Espinosa-Oviedo\inst{4}
\and Carmem Hara\inst{5}
\and Sabine Loudcher\inst{3}
\and Regina Motz\inst{7}
\and Martin A. Musicante\inst{2}
\and José-Luis Zechinelli-Martini\inst{6}
}
\authorrunning{G. Vargas-Solar et al.}
%
\institute{CNRS, Univ Lyon,  INSA Lyon, UCBL, LIRIS, UMR5205, F-69221, France\\
\email{genoveva.vargas-solar@cnrs.fr} \and
Federal University Rio Grande do Norte, DIMAP, Natal, Brazil\\
\email{\{umberto.costa,martin.musicante\}@ufrn.br}
\and 
Université Lumière Lyon 2, ERIC, France\\
\email{\{jerome.darmont,sabine.loudcher\}@univ-lyon2.fr} \and 
Université Claude Bernard Lyon 1, ERIC, France\\
\email{javier.espinosa@univ-lyon1.fr} \and
Federal University of Parana, Brazil\\
\email{carmemhara@ufpr.br} \and
Fundación Universidad de las Américas Puebla, Mexico\\
\email{joseluis.zechinelli@udlap.mx} \and
Universidad de las República, Uruguay\\
\email{regina.motz@gmail.com}
}
\maketitle              
\begin{abstract}
This paper introduces Experiversum, a lakehouse-based ecosystem that supports the curation, documentation and reproducibility of exploratory experiments. Experiversum enables structured research through iterative data cycles, while capturing metadata and collaborative decisions.
Demonstrated through case studies in Earth, Life and Political Sciences, Experiversum promotes transparent workflows and multi-perspective result interpretation. Experiversum bridges exploratory and reproducible research, encouraging accountable and robust data-driven practices across disciplines.

\end{abstract}
%
%
%
\section{Introduction}

Massive data production is increasingly vital in experimental sciences such as life, earth, social sciences and humanities, where large-scale, cost-effective data acquisition is now possible. Such fields generate diverse datasets of varying quality, enabling multifaceted analyses. Traditional schema-on-write methods such as ETL (Extraction, Transformation, Loading) struggle with such heterogeneity. Data lakes provide a flexible alternative by storing raw data in original formats, but require effective metadata extraction to integrate data and ensure reproducibility.

Open science demands not just data sharing, but also the documentation of experimental context, including conditions and decisions. This requires detailed metadata that captures both data and the knowledge production process.
The main challenge is twofold: designing metadata models that represent both data and processing workflows, and implementing ELT (Extraction, Loading, Transformation) pipelines that support experiment curation and track how decisions impact outcomes. Metadata must serve as an execution guide for ELT processes to ensure reproducibility.

This paper introduces Experiversum, a lakehouse prototype system that applies a metamodel to curate and manage data-driven experiments. Experiversum enables researchers to explore, analyse and reuse experiments with rich metadata, in alignment with open science principles.
Consequently, the remainder of the paper is structured as follows. Section~\ref{sec:related} reviews related works on metadata, provenance and reproducibility. Section~\ref{sec:process-curation} introduces the Experiversum ecosystem. Section~\ref{sec:experiversum} details the system’s architecture, curation processes and exploration functions. Section~\ref{sec:use-case} presents use cases in social, earth, and life sciences. Section~\ref{sec:conclusion} concludes and outlines future work.

\section{Related Works}\label{sec:related}
This section reviews key approaches for curating experimental data and processes, covering storage and management systems such as data warehouses, data lakes, lakehouses and dataverses \cite{paper2key,Vargas-Solar2024}. We also compares data lake solutions used in earth, life, and social sciences.

\noindent{\em The Evolving Practice of Data Curation.}
%
Data curation has evolved from focusing on preservation and quality control \cite{higgins2008dcc,lord2003science} to a value-added process that includes metadata enrichment and contextualization \cite{palmer2008purposeful}. In fields of earth sciences and biodiversity, this shift supports reusability and clear provenance \cite{cheney2009provenance}. Modern platforms such as dataverse combine automation with expert oversight to support the full research lifecycle \cite{zgolli2020metadata}.

\noindent{\em Infrastructure for Modern Research.}
%
Managing today’s research data, ranging from structured tables to unstructured content, requires flexible systems. Data warehouses are optimized for structured analytics, but lack support for diverse formats \cite{Bimonte2024}. Data lakes address this with schema-on-read flexibility \cite{Hegde2017}. However, without proper governance, data lakes risk becoming ``data swamps'' \cite{Becker2022}. The lakehouse model combines the strengths of both warehouses and lakes \cite{Armbrust2021}, while dataverses offer curated, citable storage \cite{crosas2015automating,king2007introduction}. Our work advocates for integrating a lakehouse and a dataverse approaches in earth and life sciences.

\noindent{\em Discipline-Specific Data Challenges.}
%
Different disciplines require tailored infrastructures. In natural sciences, repositories support metadata standards for reproducibility \cite{russom2016data}. In social sciences, data is often qualitative and harder to standardize. Data lakes offer needed flexibility \cite{Hai2016}, but must preserve context and consider ethical issues, especially with personal or indigenous data \cite{Carroll2020}. Hybrid solutions aim to balance scale and detail \cite{Dunning2021,Vargas-SolarDAE24}.

\noindent{\em Innovations and Remaining Challenges.}
Emerging technologies such as conversational analytics using Large Language Models (LLMs) are reshaping interaction with data\footnote{Nguyen 2024; Kerner 2023; Dubey 2024}. While promising, LLMs raise concerns about accuracy and trust\footnote{Ghodsi et al., 2023}. Interoperability remains difficult across disciplines and data types \cite{sawadogo2021data}. Ultimately, success depends not only on technical solutions but also on institutional support and user adoption \cite{Jagadish2021}.



\section{Curating Data-Driven Experiments}\label{sec:process-curation}
A data-driven experiment consists of three key elements: (1) raw data from empirical sources, (2) the research team responsible for data selection, methods, and validation, and (3) contextual metadata describing collection, processing, and analysis conditions. Curation ensures all components are documented for transparency and reproducibility.
We define a metadata model structured around three concepts: raw content, experimental specifications and context. Figure~\ref{fig:UML} in Appendix \ref{appendix} illustrates this model.
%

\noindent{\em Level 1: Raw content.} 
The blue classes in Figure~\ref{fig:UML} represent data ingested or produced during experiments. Metadata are extracted through automated and manual processes, capturing summaries, distributions and structure, e.g., column types and format. Each release is profiled, e.g., licensing, size and provenance; and can include tabular, textual or signal data. Items can also be annotated with multimedia or textual comments to enhance interpretability. 
%

\noindent{\em Level 2: Experimental specifications.}
This level documents actions performed on datasets, whether manual or automated. Actions produce artefacts or models, with metadata describing structure, execution and provenance. Parameters, evaluation criteria and validation protocols are recorded to trace how and why actions were performed or repeated.

\noindent{\em Level 3: Experiment context.}
Metadata describe the research team’s composition , e.g., roles and seniority; responsibilities and the guiding research question. It captures the decision-making context and provides a basis for comparing experiments.

\section{Experiversum: experiments and data universum environment} \label{sec:experiversum}
\begin{figure}[h]
\centering
\includegraphics[width=0.98\linewidth]{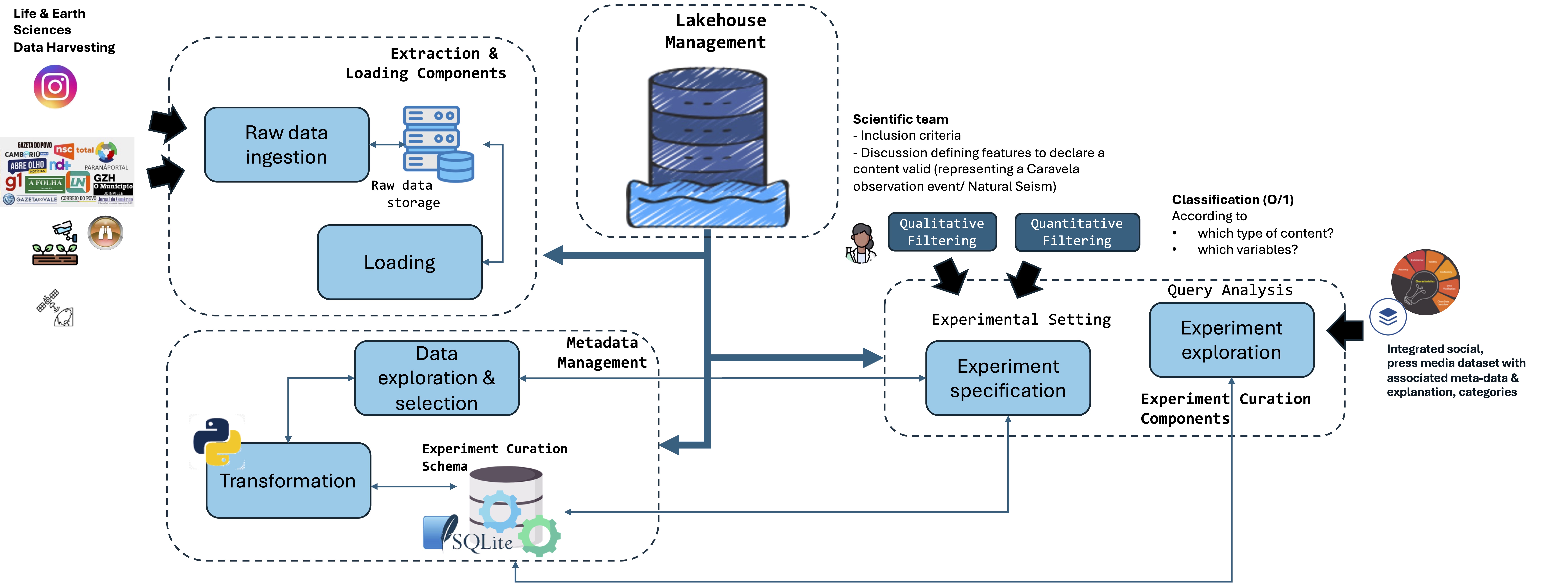}
\caption{Experiversum Architecture}
\label{fig:architecture}
\end{figure}

The Experiversum environment ensures preservation, documentation and reproducibility of scientific experiments using a lakehouse infrastructure (Figure~\ref{fig:architecture}).

\noindent{\em Extraction and Loading.}
Raw data such as seismic signals or social media posts are ingested in their original format, e.g., signals, text and media. Subsets are grouped into catalogues based on attributes such as ingestion date, size, format and quantitative traits.

\noindent{\em Metadata Management.}
This module links metadata from raw and processed data to the experimental context, supporting reproducibility. Scientists can navigate datasets using quantitative summaries and relevant descriptors. Metadata is extracted following the curation model, normalized, and stored in a metadata repository, enabling comparison and exploration across experiments.

\noindent{\em Experiment Curation.}
%
Researchers can specify experimental parameters such as selection criteria, team roles, questions and performance constraints. They can explore experiments on similar topics conducted under different conditions, review methods and assess outcomes—supporting comparative analysis and enhancing reproducibility.

\noindent{\bf Experiversum management.} 
This component orchestrates Experiversum pip-elines, managing seamless data flow from ingestion to analysis. It ensures consistency, performance, and smooth transitions across the infrastructure.

\noindent{\em Extraction and Loading Pipeline.}
The EL pipeline handles data ingestion, cleaning and transformation before loading it for analysis—crucial for any data workflow.
Key steps include (i) data extraction retrieves raw data from sources such as APIs, sensors, databases or unstructured files, e.g., seismic logs and social media; (ii) data cleaning removes duplicates, errors and inconsistencies; (iii) data enrichment  adds contextual details such as timestamps or geolocations reliability; (iv) data loading stores data with associated structural and quantitative metadata in organized collections.

\noindent{\em Tagging Experimental Processes Pipeline.}
%
improves reproducibility and collaboration by assigning structured tags to experimental workflows. It consists of three tasks. (i) experiment specification records metadata such as experiment ID, name  and date to link processes and data; (ii tagging applies algorithmic or user-defined tags to annotate datasets and processes; (iii) tag storage maintains tag traceability for reuse and reference.

\noindent{\em Transformation Pipeline}
 converts raw or semi-structured data into usable formats aligned with the metadata model. It consists of three tasks. (i) structuring maps text, signals and media to metadata entities; (ii) contextual enrichment adds metadata to reflect experimental settings; (iii) preparation formats data for analytics, machine learning or further experimentation.

\noindent{\bf Exploring and Querying Processes.}
The exploration and analytics pipeline enables users to query, analyze and visualize curated datasets. It supports exploratory data analysis (EDA), statistical modeling and machine learning to uncover insights. There are five tasks.

\noindent - Experiment Querying and Retrieval: access datasets by filtering parameters such as time, location or experiment settings for efficient data selection.\\
\noindent - Filtering and Aggregation: refine data by extracting relevant subsets and aggregating across dimensions, e.g., time and  region) to produce summary metrics.\\
\noindent - Descriptive and Predictive Analytics: perform statistical analysis (averages, correlations, trends) and advanced tasks (classification, regression, clustering, anomaly detection...) for pattern discovery and forecasting.\\
\noindent - Data Visualisation: display results using graphs, charts and heatmaps to simplify interpretation, trend spotting and anomaly identification. \\
\noindent - Collaboration and Sharing: share results, export outputs and integrate findings into reports or publications to support teamwork and dissemination.

\section{Use Case-Based Validation}\label{sec:use-case}

The first prototype of Experiversum is implemented using SQLite3 as the storage backend. Pipelines are developed in Python and three demonstration scenarios (biodiversity, seismic data and graffiti analysis) are built using Flask, Bootstrap and executable Jupyter notebooks.

\noindent{\bf Tracking ``Caravelas Portuguesas'' along the Brazilian Coast.}
This use case classifies sightings of the jellyfish Physalia physalis along Brazil’s coast\footnote{\url{https://es.wikipedia.org/wiki/Physalia\_physalis}}.

\noindent\textit{Raw data extraction and loading.}
Instagram posts tagged with relevant hashtags (\#aguaviva, \#caravelaportuguesa, etc.) are extracted, converted into CSVs containing metadata (ID, source, location, media URL) and uploaded into Experiversum.

\noindent\textit{Data transformation.}
CSV headers are mapped to our metadata model, e.g., experiment, media, content and tags). Unstructured and imprecise geo-temporal data, e.g., ``last summer'' or inaccurate locations, is cleaned and corrected. Each transformation is registered and results in derived datasets.

\noindent\textit{Experimental settings.}
Two research teams collaborate: data scientists use machine learning models to classify posts, while biologists manually tag and define classification categories. Settings include inclusion criteria, e.g., location/time) gender of the affected person, model calibration and performance thresholds.

\noindent\textit{Exploration and querying.}
Users query jellyfish occurrences by time and region, explore ecological associations and compare human and machine learning classifications to study methodological differences.

\noindent{\bf Classification of Seismic Activity in Northeast Brazil.}
This case curates seismic data to differentiate natural from anthropogenic events, producing validated bulletins summarising seismic activity.

\noindent\textit{Experiment setup.}
Participants include seismographs (data generation), data collectors (retrieval), junior analysts (event detection) and senior analysts (review and bulletin publication).

\noindent\textit{Data extraction and loading.}
SAC files are uploaded and validated. Amplitude values (by axis: X, Y, Z) are extracted and stored along with metadata such as  station\_id, channel\_id and timestamps.

\noindent\textit{Data transformation and tagging.}
Junior analysts plot waveforms to detect and tag events. Each event is annotated by station, year and magnitude. Analysts identify P and S wave arrivals. A triangulated event list forms the basis of the official bulletin, validated by a senior analyst.

\noindent\textit{Exploration.}
Waveforms and results are visualised through a Web interface. Analysts can share or publish curated outcomes.

\noindent{\bf Graffiti Analysis for Political Messaging.}
A two-member team (junior + senior) conduct qualitative analysis to classify political graffiti across a city.

\noindent\textit{Research framing.}
Over two cycles, the team refines the central question: ``Can political messages be traced through graffiti?'' They define inclusion criteria and political graffiti indicators through discussion.

\noindent\textit{Data collection.}
The junior researcher photographed 1,050 graffiti images across districts. After review, 546 were validated and shared on Instagram (link upon acceptance).

\noindent\textit{Analysis.}
Manual classification is complemented by unsupervised machine learning  (k-means and hierarchical clustering via Orange). Results from both are iteratively refined and interpreted collaboratively.

\noindent\textit{Results.}
Narratives and metadata are compiled through successive review rounds. Final deliverables include classifications, summaries and reproducibility documentation.

\noindent{\bf Lessons Learned.} 
%
Developing an experiment curation system reveals key insights into the challenges and benefits of structuring data-driven research. It underscores how data, metadata and decisions intersect, and the importance of systematic curation for transparency, reproducibility and collaboration.

\noindent{\em Curation and Reproducibility.}
Experiversum supports the curation of varied data types (seismic signals, social media and multimedia) through ingestion, transformation and tagging pipelines. These pipelines enrich content with contextual metadata, enabling reproducible experiments and traceable results.\
\textit{Lesson}: Metadata models are crucial for linking data with experiments, ensuring interpretability beyond storage.

\noindent{\em Data Transformation and Tagging.}
While structured data such as seismic signals are easily processed, tagging unstructured content, e.g., social media, prove more difficult, requiring advanced techniques.\
\textit{Lesson}: Automated tagging suits structured data. Unstructured sources need robust Natural Language Processing (NLP) methods.

\begin{figure}
\centering
\includegraphics[width=0.9\linewidth]{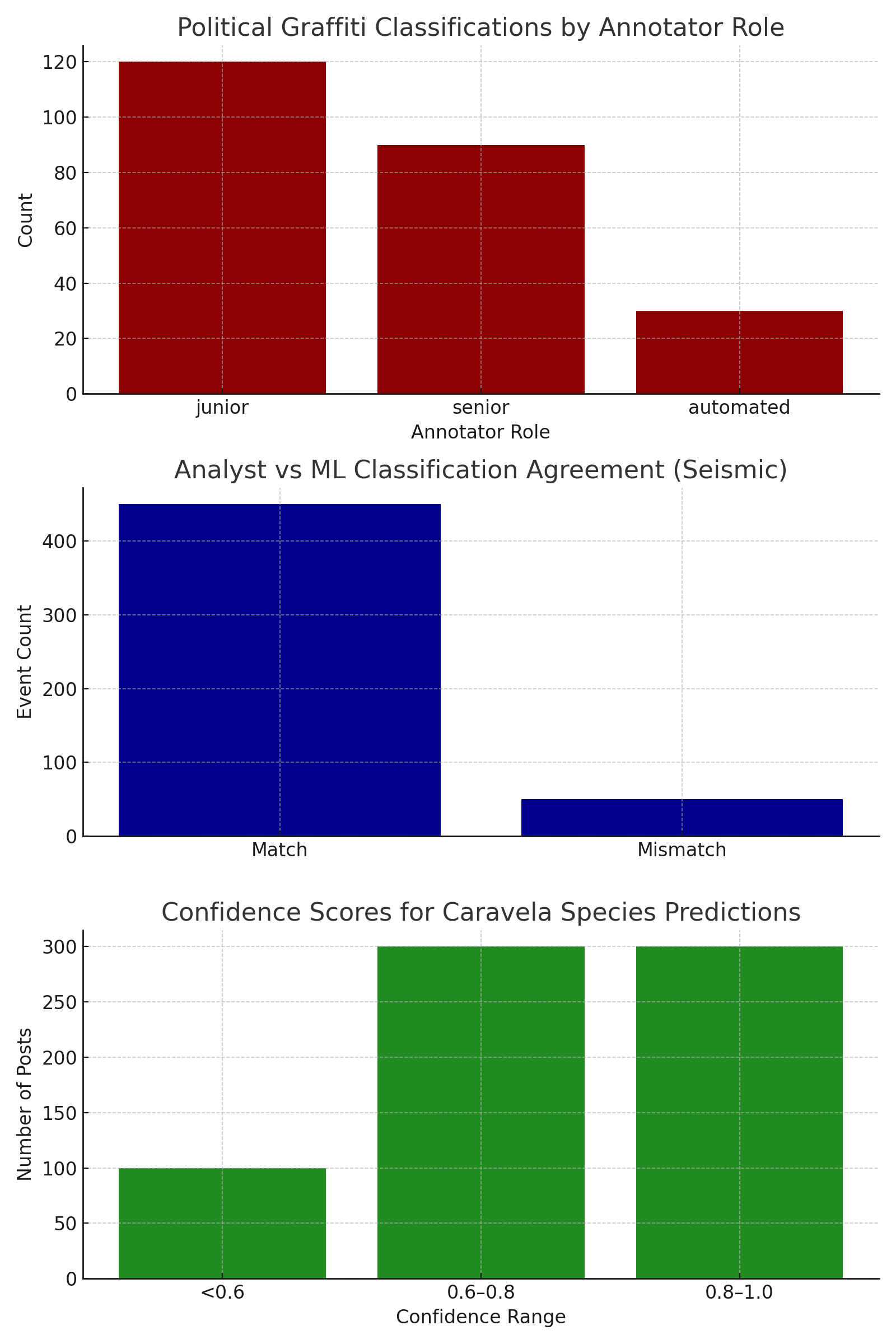}
\caption{Query visualisation in Experiversum}
\label{fig:experiments}
\end{figure}

\noindent{\em Using Metadata to Understand Experiments.}
Figure~\ref{fig:experiments} illustrates metadata-driven queries across three use cases. The first chart shows political graffiti labels by annotator, with juniors contributing most tags, indicating their key role in interpretation. The second chart compares human and machine classifications in seismic monitoring, showing 90\% agreement but highlighting some discrepancies needing expert review. The third chart displays confidence scores for species classification, while many fall in the 0.8–1.0 range, lower-confidence cases ($<$0.6) point to the need for manual checks. Such visualisations show how curated metadata improves analysis, validation and understanding across complex experiments.

\section{Conclusion and Future Work}\label{sec:conclusion}

This paper introduces Experiversum, a lakehouse-based platform for curating, exploring and reproducing data-driven experiments. Experiversum integrates ELT pipelines with a structured metamodel to link raw data to experimental intent, enabling workflow reuse across disciplines. Case studies with biodiversity and seismic data highlight its flexibility for interdisciplinary research.
The main insight is that reproducibility requires preserving full experimental context, not just raw data. Our metadata model and curated workflows improve traceability and reuse across diverse data types.

Future work includes extending metadata coverage, using NLP and graph techniques for tagging, adding privacy-aware analytics, and deploying the platform in real infrastructures to support open, collaborative science.

\subsubsection{Acknowledgements.} This work was funded by project LETITIA, Lyon  Computer Science Federation (FIL). {\url{http://www.vargas-solar/letitia}}

%
%
%
 \bibliographystyle{splncs04}
 \bibliography{biblio}

\newpage
\appendix
\section{Appendix A}\label{appendix}
\begin{figure}[h]
\centering
\includegraphics[width=0.95\linewidth]{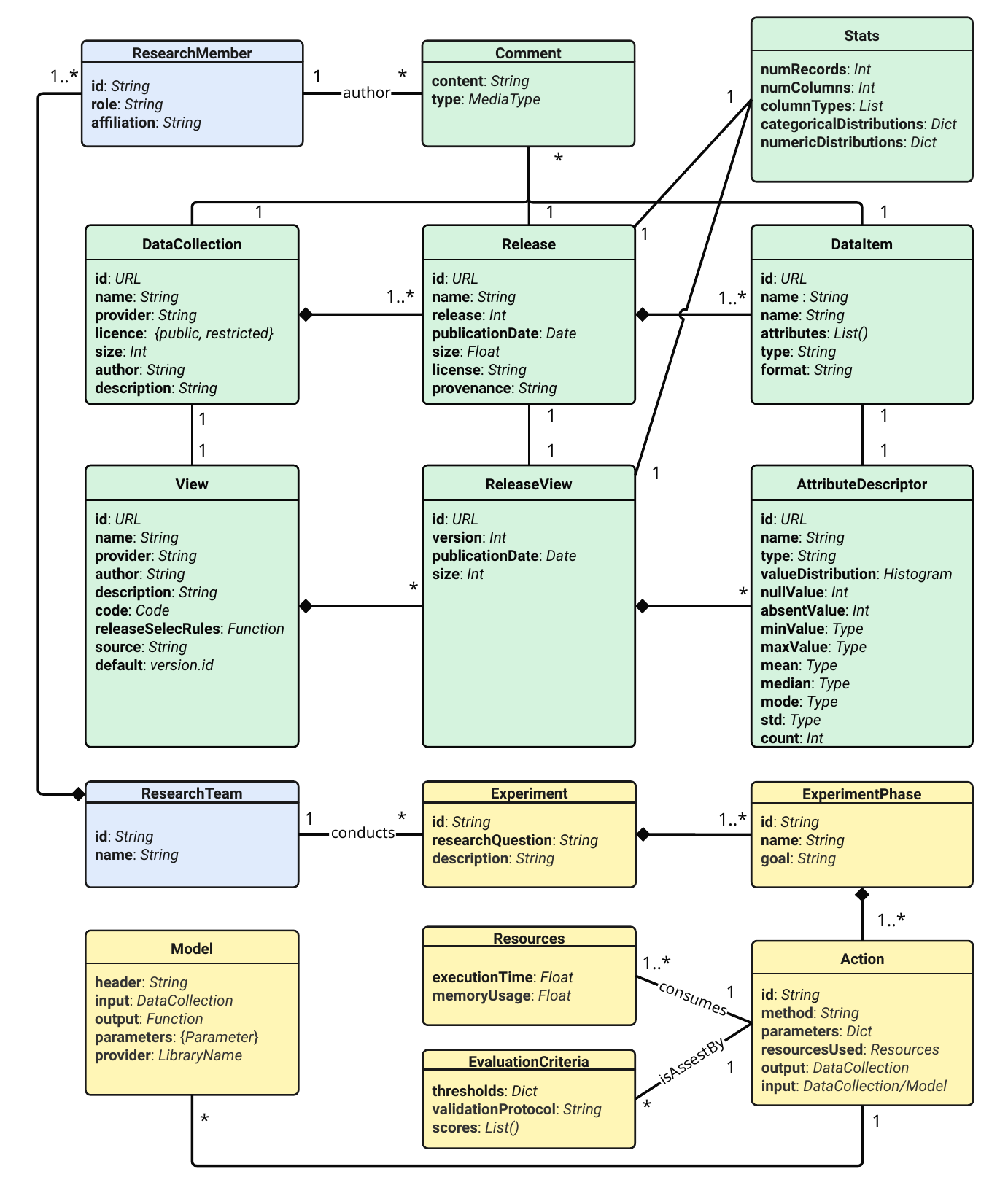}
\caption{Data Metamodel UML Class Diagram}
\label{fig:UML}
\end{figure}

\end{document}